\documentclass[conference]{IEEEtran}
\IEEEoverridecommandlockouts
\usepackage{cite}
\usepackage{amsmath,amssymb,amsfonts}
\usepackage{algorithmic}
\usepackage{graphicx}
\usepackage{textcomp}
\usepackage{xcolor}
\usepackage{booktabs}
\def\BibTeX{{\rm B\kern-.05em{\sc i\kern-.025em b}\kern-.08em
    T\kern-.1667em\lower.7ex\hbox{E}\kern-.125emX}}

\makeatletter
\newcommand{\linebreakand}{%
  \end{@IEEEauthorhalign}
  \hfill\mbox{}\par
  \mbox{}\hfill\begin{@IEEEauthorhalign}
}
\makeatother

\begin{document}

\title{A Hybrid Method of Sentiment Analysis and Machine Learning Algorithm for the U.S. Presidential Election Forecasting}

\author{\IEEEauthorblockN{Guocheng Feng}
\IEEEauthorblockA{\textit{Guangdong Provincial Key Laboratory of Interdisciplinary} \\
\textit{Research and Application for Data Science} \\
\textit{BNU-HKBU United International College}\\
Zhuhai, China \\
q030026233@mail.uic.edu.cn}
\and
\IEEEauthorblockN{Kaihao Chen}
\IEEEauthorblockA{\textit{Department of Statistics and Data Science} \\
\textit{BNU-HKBU United International College}\\
Zhuhai, China \\
q030026012@mail.uic.edu.cn}
\linebreakand
\IEEEauthorblockN{Huaiyu Cai}
\IEEEauthorblockA{\textit{Department of Statistics and Data Science} \\
\textit{BNU-HKBU United International College}\\
Zhuhai, China \\
q030026003@mail.uic.edu.cn}
\and 
\IEEEauthorblockN{Zhijian Li}
\IEEEauthorblockA{\textit{Guangdong Provincial Key Laboratory of Interdisciplinary} \\
\textit{Research and Application for Data Science} \\
\textit{BNU-HKBU United International College}\\
Zhuhai, China \\
zhijianli@uic.edu.cn}
}

\maketitle

\begin{abstract}
U.S. Presidential Election forecasting has been a research interest for several decades. Currently, election prediction consists of two main approaches: traditional models that incorporate economic data and poll surveys, and models that leverage X and other social media platforms due to their increasing popularity in the past decade. However, traditional approaches have predominantly focused on national-level predictions, while social media-based approaches often oversimplify the nuanced differences between online discourse and the broader voting population's political landscape.

In this work, we perform a hybrid method of both the machine learning algorithm and the sentiment analysis on the state level with various independent variables including census data, economic indicators, polling averages, and the newly defined average sentiment scores from X.
Our prediction for the 2020 U.S. Presidential Election yielded promising results. Most of our models successfully predicted a victory for the Democratic candidate with 96\% accuracy using Gradient Boosting Trees and Multi-Layer Perceptron algorithms. This novel prediction framework addresses the limitations of existing U.S. Presidential Election forecasting approaches, particularly in terms of state-level predictions. It provides a valuable foundation for future research in this field and contributes to advancing our understanding of election dynamics.
\end{abstract}


\section{Introduction}
The United States (U.S.) Presidential Elections forecasting has long been a research interest among statisticians, and politicians for decades \cite{fair2011predicting}, and has also been attempted by data scientists since the democratization of information in the digital platform era. Despite considerable efforts made by scholars in this field, the U.S. Presidential Election is still considered a difficult task for many reasons. 

Since the 1980s, research in the field of predicting the U.S. Presidential Elections has largely focused on equation-based models that incorporate political-economic theories at the national level \cite{LewisBeck2014USPE}, as well as the direct or indirect use of polling data at both the national and state levels. Notable examples of such models include Nate Silver's work \cite{silver2012signal} that successfully predicted the outcomes of the 2012 election in all 50 states, Drew Linzer's model - the Votamatic \cite{linzer2013dynamic}, and Simon Jackman's Uniform Swing model \cite{jackman2014predictive}.

However, in recent years, as presidential elections have grown progressively competitive (such as the 2016 election), these methods have generally failed to accurately predict the final outcome of the presidency. While many have successfully predicted that Hillary Clinton would win the popular vote with a small margin compared to actual results \cite{Campbell2017ARO}, state-level prediction results thereof are generally inaccurate and thus failed to predict the electoral votes. Therefore, when predicting the U.S. Presidential Elections, high accuracy at the national level might not be sufficient for correct forecasting, and poll surveys might suffer from sampling biases \cite{Valentino2017PollingAP}. Thus, we want to design an election prediction framework that takes into account both state-level opinion dynamics and robust theoretical approaches from political science.

Additionally, social media has been an important platform for political discourse, leading researchers to examine the potential for predicting presidential election outcomes using data from platforms such as Facebook and X (formerly known as Twitter). The predominant approaches in this field have focused on sentiment analysis, topic modeling, and social network analysis, often utilizing tweets (posts made on X) due to their accessibility. Current literature on the predictive power of tweets predominantly focuses on substituting traditional polling with tweet sentiments \cite{OConnor2010FromTT, Beauchamp2017PredictingAI} or developing indicators based on tweet sentiments \cite{Yavari2022ElectionPB}. However, the use of tweet data presents certain limitations, such as X users tend to be not representative of the U.S. population \cite{Mislove2011UnderstandingTD}, and tweets data exhibit higher bias in popular vote compared to traditional polls \cite{Anuta2017ElectionBC}. As such, solely utilizing the X platform may not be comprehensive enough for election prediction, highlighting the need for data from other sources.

To address the limitations of existing approaches, our research aims to develop an integrative model that incorporates both macro-level dynamics, such as demographic and economic data, and micro-level dynamics, including opinions mined from polling and social media data. By combining these various data sources from all 50 states, as well as the District of Columbia, we seek to enhance the accuracy of our predictions. To our best knowledge, although Liu et al. \cite{liu2021can} demonstrate a prediction method at the county level using economic data augmented with sentiment scores, the predictive power thereof in national presidential elections remains an underexplored topic.

To fill the gap, our approach begins by leveraging an election prediction framework inspired by political science, which incorporates economic and polling data. To capture the influences of late-breaking events and mitigate potential sampling biases of traditional polls, we augment our model with tweets sentiment analysis. Furthermore, to account for the local demographic estimate of voting preferences, we incorporate census information into our model. This enables us to capture the evolving composition of the electorate and its potential impact on election outcomes. By utilizing these predictive indicators and analyzing the relationship between them and election outcomes, we hope to provide policymakers, researchers, media professionals, and the public with a more reliable and informative prediction framework for future presidential elections.

\begin{figure}
\centerline{\includegraphics[width=0.47\textwidth]{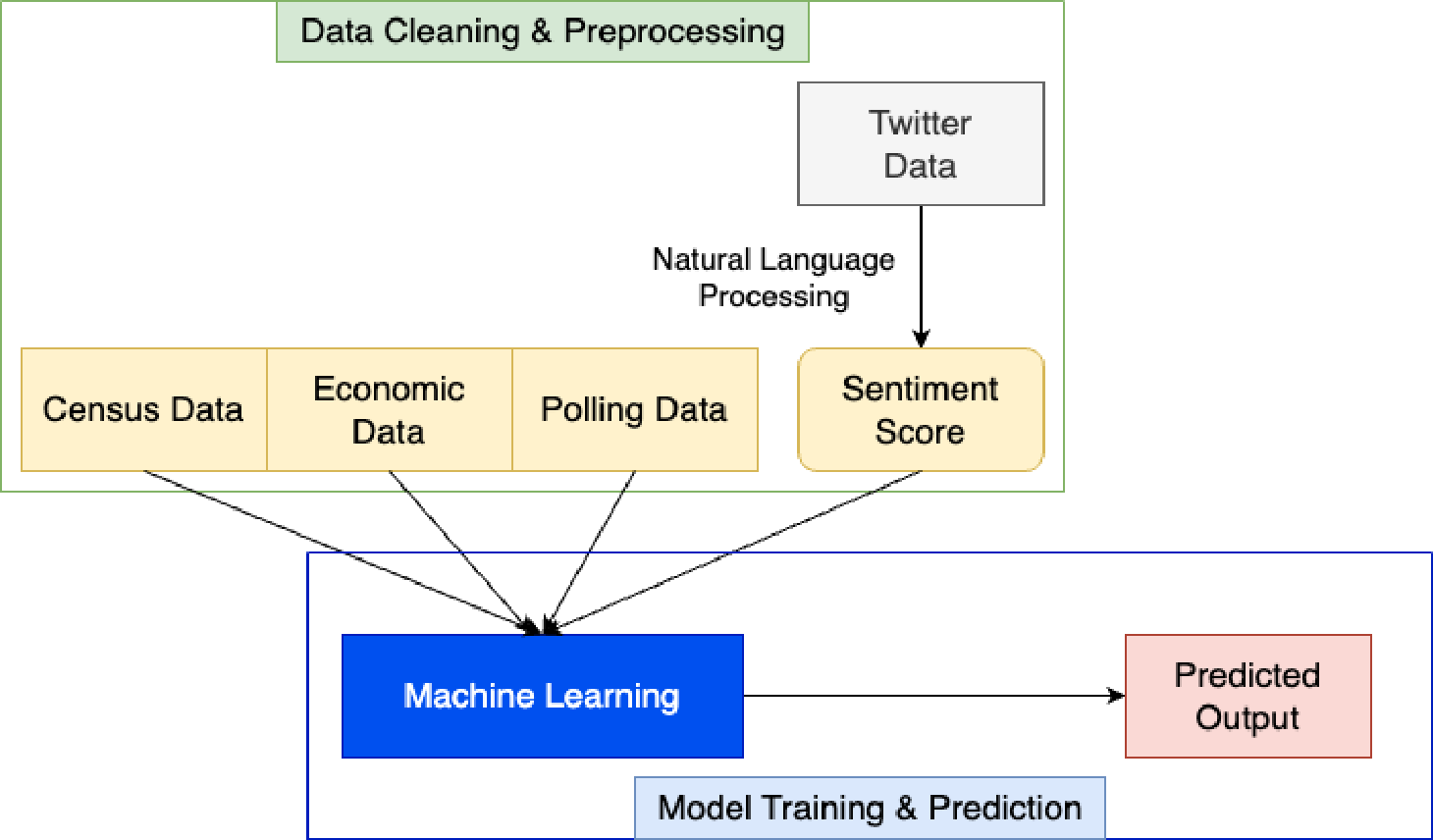}}
\caption{The Overall Structure of Election Prediction}
\label{workflow}
\end{figure}

\section{Methodology}

\subsection{Data}
The data used in our election prediction model consists of four primary types: census data, economic data, the 2012, 2016, and 2020 presidential election polling data, and tweets related to presidential candidates. A detailed list of variables selected for model predictions is provided in Table \ref{datafeatures}. Detail information of our data source can be accessed via our Github repository\footnote{https://github.com/matt-feng/US\_Election\_Prediction}.

\subsubsection{Census Data}
Nickerson et al. \cite{CensusReference} provide a holistic analysis on leveraging census data to develop election indicators. In addition to their work, we add ethnic data to account for associations between ethnic groups and their vote preferences \cite{houle2019religion}.

\subsubsection{Economic Data}
Economic data includes variables that reveal regional economic conditions. Lewis-Beck and Stegmaier \cite{lewis2000economic} emphasize the strong effects of GDP and personal income on government votes. Thus, we select the two variables following their insights into the relationship between economics and elections.

\subsubsection{Polling Data}
The polling data is a traditional method to mine people's opinions which indicates the support rate of the presidential candidates and directly reflects people's preference before the final election day. We use the average of polling survey results corresponding to each state from August to election day in 2012, 2016, and 2020 to smooth out the fluctuations.

\begin{table}
\centering
\caption{Variables Included in the U.S. Presidential Prediction Dataset}
\begin{tabular}{lp{5.5cm}}
\toprule
\textbf{Type} & \textbf{Variables}\\
\midrule
Census Data &  Age Distribution, Family Income, Occupations Distribution, Poverty, Race, Sex Ratio, Unemployment Rate\\
\midrule          
Economic Data & GDP, Per capita personal income\\
\midrule
Election Polls & Democrat and Republican Polling Average\\
\midrule
Tweets Sentiments & Democrat and Republican Sentiment Scores Average (Positive and Negative)\\

\bottomrule
\end{tabular}
\label{datafeatures}
\end{table}

\subsubsection{Tweets Data}
We use SNSCRAPE\footnote{https://github.com/JustAnotherArchivist/snscrape}, a widely-used web scraper tool for scanning social networking services, to collect tweets in English related to presidential candidates with selected keywords in Table \ref{tweetkeys}. 
We then use the hashtag tracking tool Hashtagify\footnote{https://hashtagify.me}, a website that analyzes trends in hashtag usage within tweets and that has been used in many studies \cite{TurnerMcGrievy2015TweetFH, Harris2014ArePH}, to confirm the selected keywords are commonly used in the context of presidential elections. The dataset includes tweets posted from the first presidential election debate of each election year up to the election day. To conduct state-level dynamic analyses, we extract tweets that were tagged with location information and classify them according to their corresponding state or district. Specifically, we classify the tweets into 50 states plus the District of Columbia, allowing for an in-depth examination of state-level trends and patterns in the data.

\begin{table*}
\centering
\caption{Keywords for filtering tweets}
\begin{tabular}{lllp{8.5cm}}
\toprule
\textbf{Year} & \textbf{Time Range} & \textbf{Candidates} & \textbf{Keywords}\\
\midrule
2012 & Oct. 3 - Nov. 6 & Obama  & barackobama, obama, obama2012, democratic, democrat\\
     &                & Romney & mittromney, romney, mitt2012, republican, gop\\
\midrule
2016 & Sept. 26 - Nov. 8 & Clinton & hillaryclinton, clinton, clinton2016, demoratic, democrat\\
     &                & Trump & realdonaldtrump, donaldtrump, trump, trump2016, maga, republican, gop\\
\midrule
2020 & Sept. 29 - Nov. 3 & Biden & joebiden, biden, biden2020, demoratic, democrat\\
     &                & Trump & realdonaldtrump, donaldtrump, trump, trump2020, maga, republican, gop\\
\bottomrule
\end{tabular}
\label{tweetkeys}
\end{table*}

\subsection{Sentiment Analysis of Tweets}
After extracting data from X, our dataset comprised a total of \textbf{1,731,554} unique tweets and \textbf{392,718} identified users. To perform the tweet data preprocessing, we remove hyperlinks, hashtag symbols (\#), and mentions (@USER) as they are generally irrelevant to our textual analysis. Next, we utilize a pre-trained Robustly optimized BERT approach (RoBERTa) model \cite{loureiro-etal-2022-timelms} to detect the presence of positive, neutral, and negative sentiments within tweets. RoBERTa is trained with larger datasets and demonstrates improved performance compared to traditional lexicon-based models \cite{Liu2019RoBERTaAR}.

To evaluate the performance of our sentiment analysis model, we conducted an experiment using four popular pre-trained transformer models including RoBERTa, XML, BERT, and xDistil. We manually annotate 150 sample tweets, assigning each tweet a single sentiment label. Next, we apply the models to predict the sentiment for each tweet, assigning the label with the highest score as the predicted outcome. The evaluation results are summarized in Table \ref{sent_eval}. Among the models tested, RoBERTa achieved the best scores across all evaluation metrics, with an accuracy of 68\%. However, the F1-score of our RoBERTa models does not guarantee accurate classification, and we leave for future research to employ more advanced methods for accuracy improvement since sentiment analysis is not the scientific contribution of this study.

\begin{table}
\centering
\caption{Sentiment Analysis Performance of Four Models}
\begin{tabular}{lcccc}
\toprule
\textbf{Classification Model} & \textbf{Accuracy} & \textbf{Precision} & \textbf{Recall} & \textbf{F1}\\
\midrule
RoBERTa & \textbf{0.68} & 0.74 & 0.63 & 0.63\\
XML     & 0.61 & 0.63 & 0.61 & 0.59\\
BERT    & 0.64 & 0.61 & 0.62 & 0.61\\
xDistil & 0.61 & 0.60 & 0.47 & 0.48\\
\bottomrule
\end{tabular}
\label{sent_eval}
\end{table}

\subsection{Define the average sentiment score as a new feature}
\begin{figure}
\centerline{\includegraphics[width=0.48\textwidth]{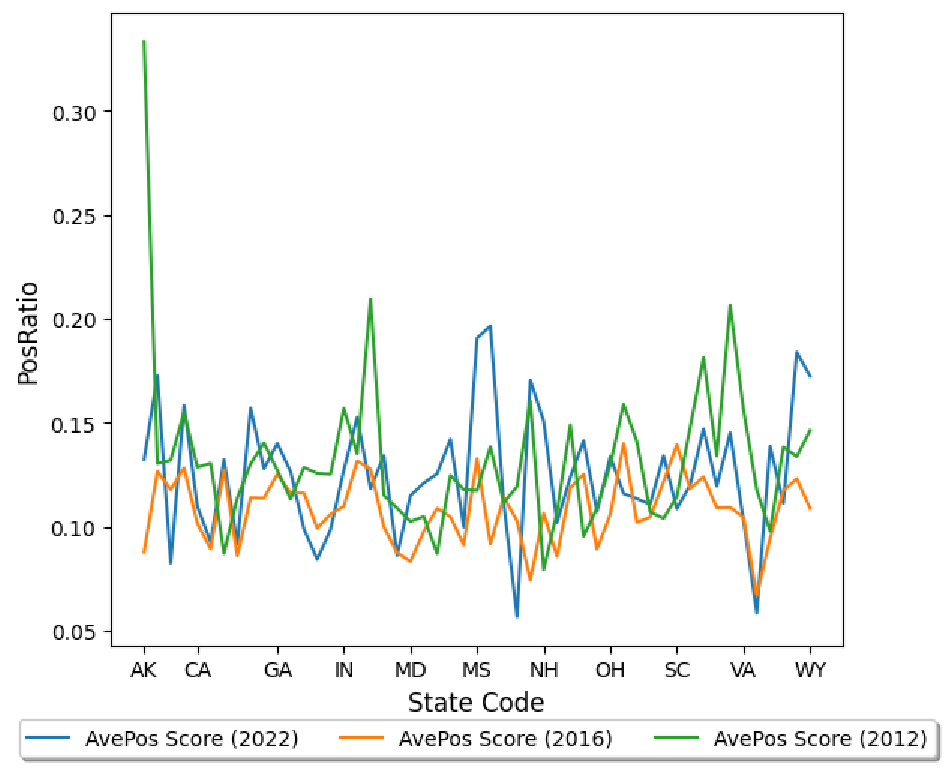}}
\caption{The average positive scores of the Republican Party in 2012, 2016, and 2022}
\label{sent_2020}
\end{figure}

For every tweet, the RoBERTa model produces three probability scores: positive (pos), neutral (neu), and negative (neg). These probabilities represent the likelihood of each sentiment category, with the constraint that the three probabilities sum up to 1. To assign a sentiment label to each tweet, we select the category with the highest probability. Next, we define the average positive sentiment score $AP.score$ and the average negative sentiment score $AN.score$ pertaining to a candidate within a specific state. Both $AP.score$ and $AN.score$ are computed as the proportion of positive or negative tweets with respect to the total number of state-related tweets concerning the candidate. Thereby, the average neutral sentiment score $AT.score$ can be calculated as $AT.score = 1- AP.score - AN.score$. For illustrative purposes, we present in Figure \ref{sent_2020} the $AP.score$ values for the Republican party across all 50 states, along with the District of Columbia, during the 2012, 2016, and 2020 elections. The $AP.scores$ for the Republican Party demonstrate consistent patterns across all three elections. However, there is an unusual peak in Alaska during the 2012 election, and we suspect that the lack of data may have contributed to this situation.

It is worth mentioning that instead of using positive/negative tweet counts to measure the support rate for a presidential candidate, our defined average sentiment score can essentially avoid the uneven number problem of tweets from two candidates.

\subsection{Combining sentiment analysis and machine learning algorithm}
Next, we apply the machine-learning algorithm to do classification and prediction with our newly defined features the average sentiment score $AP.score_{ij}$ and $AN.score_{ij}$, and features from census data, economic data, and polling data.

\section{Results}
To essentially evaluate our approach and also for comparison purposes, we apply seven classification models for election forecasting. Our dataset contains 2012, 2016, and 2020 election data. The binary response is the voting result of a candidate in each state. The independent variables are as mentioned above, including the average sentiment scores, census data, economic data, and polling data. For our case study, we conduct a 10-fold cross-validation on the entire dataset utilizing the 2020 U.S. presidential election data for validation, and the 2012 and 2016 election data for training.

Model results of a 10-fold cross-validation are shown in Table \ref{eval}. Ensemble methods, such as the Gradient Boosting Trees and the random forest have the highest accuracy. The result may be due to the limitation of the small dataset since we only have 153 data samples in total where each state in a year in 2012, 2016, or 2020, is a sample. While ensemble methods can reduce the risk of overfitting, methods like Support Vector Machine and Gaussian Naive Bayes are more sensitive to noises in data and thus generate sub-optimal classification accuracy.

\begin{table}
\centering
\caption{Evaluation of Seven Classification Models using 10-Fold Cross-Validation}
\begin{tabular}{lcc}
\toprule
\textbf{Classification Models} & \textbf{Accuracy} & \textbf{Accuracy (+sentiments)}\\
\midrule
Gradient Boosting Trees & 95.4\% & 95.4\%\\
Decision Tree & 90\% & 93\%\\
Logistic Regression & 91.5\% & 92.8\%\\
Support Vector Machine & 87.5\% & 90.2\%\\
Random Forest & 92.7\% & 95.3\%\\
Gaussian Naive Bayes & 86.2\% & 88.1\%\\
Multi-Layer Perceptron & 91.4\% & 94\%\\
\bottomrule
\end{tabular}
\label{eval}
\end{table}

As we expected, Table \ref{eval} shows that incorporating our tweets sentiment scores feature in the model will noticeably enhance the overall accuracy. Similar to previous findings \cite{shwartz2022tabular}, gradient boosting trees has relatively better performance on tabular data. However, our results shows that introducing tweets sentiment scores does not improve accuracy thereof and a more detailed analysis would be provided in 2020 U.S. Presidential Election Prediction task. Additionally, it is noticeable that our dataset comprises a diverse range of data attributes and a relatively small number of samples, which may cause the risk of over-fitting in prediction models. Therefore, we identify the top 10 most important features from tree models, as shown in Table \ref{tree importance}. 

The two polling averages majorly contribute to the models, indicating the selection and processing of polls are significant to our election prediction. Also, it is worth noting that the other three kinds of data also have moderate to strong importance to the models, specifically occupation distribution in census data. This indicates that the incorporation of census, economic, and tweets data has the potential to predict elections in the future.

\begin{table}
\centering
\caption{Feature Importance of Decision Tree Models}
\begin{tabular}{lc}
\toprule
\textbf{Feature} & \textbf{Importance}\\
\midrule
Polling\_Democrat(\%) & 0.215 \\
Polling\_Republican(\%) & 0.187 \\
Management, business, science, and arts occupations & 0.090 \\
Production, transportation, and material moving occupations & 0.080 \\
Median Family Income & 0.074\\
$AN.score_{Democrat}$	& 0.05 \\
Natural resources, construction, and maintenance occupations & 0.048 \\
$AP.score_{Democrat}$	& 0.043 \\
35-59 years (\%) & 0.034 \\
Below Poverty (\%) & 0.023 \\
PI\_Q3 & 0.019 \\
\bottomrule
\end{tabular}
\label{tree importance}
\end{table}

While our models may exhibit high accuracy levels of over 90\%, successful prediction of the final presidency is still challenging. This is particularly true for recent U.S. Presidential Elections, which largely depend on a small number of pivotal ``swing states."  To refine our approach and address this challenge, we will use the 2020 U.S. election as a case study to demonstrate the complexities involved in election prediction and possible corresponding solutions.

\subsection{2020 U.S. Presidential Election Prediction}

To predict the outcome of the 2020 U.S. Presidential Election, we trained models using election data from 2012 and 2016 as the training set. However, comparing to previous results, the accuracy of models trained solely on 2012 and 2016 data was generally lower. 

This discrepancy was particularly evident in swing states. For instance, both Arizona and Georgia had been won by Republicans in the 2012 and 2016 U.S. Presidential Elections but shifted to Democrats in 2020. Furthermore, accurately capturing the political dynamics of states like Michigan, Pennsylvania, and Wisconsin proved challenging for the models. In 2016, Donald Trump won these states by a narrow margin, and factors such as slow-moving census and economic data, as well as polling and tweets sentiments favoring Democrats, posed difficulties for accurately predicting their outcomes. 

To address these challenges that our model may not capture the changing dynamics and shifts in voter behavior that occurred in above mentioned states, we use threshold tuning on the prediction process to adjust the model behavior and thus render more flexibility for model prediction.

\begin{figure}
\centerline{\includegraphics[width=0.46\textwidth]{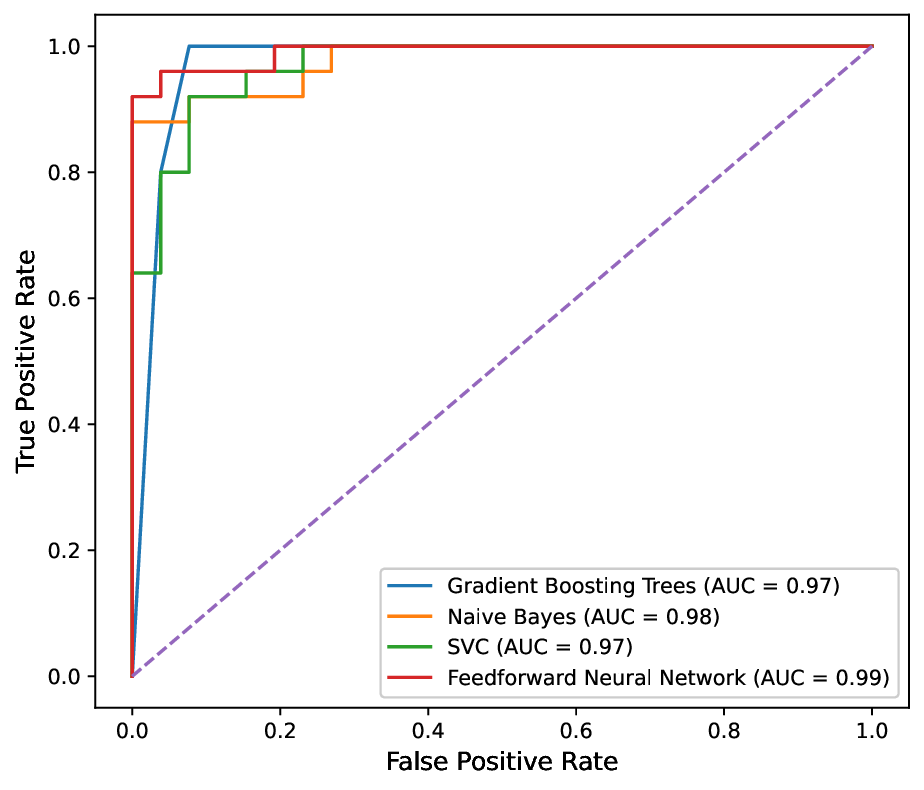}}
\caption{The ROC Curve of Models}
\label{ROC}
\end{figure}

Fig. \ref{ROC} displays the ROC curve for four classification models, in which the true positive rate (TPR) measures the accuracy of predicting states won by Republicans. We observed that when the TPR approaches 0.9, the false positive rate (FPR) remains relatively low. However, for TPR values exceeding 0.9, the FPR significantly increases. Thus, we set the prediction threshold to 0.9 across all the classification models. 

Performances of classification models are shown in Table \ref{2020eval} (model accuracy is averaged value of 10 times), in which the Voting Classifier is an ensemble method that combines the other six classification models into a single combined classifier.  All classification models successfully predict that Democrats ultimately win the 2020 U.S. Presidential Election. The Gradient Boosting Trees model and Multi-Layer Perceptron (MLP) have the highest accuracy of 96\%, which suggests that only two states were predicted to be wrong. Following is the Voting Classifier, which achieves an accuracy of 94\%. Detailed classification reports of two different threshold levels using the XGBoost algorithm implemented by Gradient Boosting Trees and MLP are shown in Table \ref{XGB&FNN}. We notice that increasing the threshold level for predicting states won by Republicans, rather than using a fixed threshold of 0.5, significantly improves the overall accuracy. In other words, rather than relying solely on the trends observed in previous elections (such as the cases of Arizona and Georgia), or directly adopting the situations that occurred in the 2016 U.S. Presidential Elections (where significant discrepancies existed between sentiment analysis and polling data, and the actual election outcomes), we propose a more nuanced approach. By setting a threshold for model predictions, we enhance the ability thereof in understanding the contextual factors that may impact election outcomes. 

Concerning the prediction results in terms of states, XGBoost model and MLP model made different mistakes. The XGBoost model incorrectly classified Arizona and Georgia as Republicans, while the MLP model misclassified Alaska and Florida as Democrats. Interestingly, when we compared the classification results with and without incorporating sentiment scores from tweets, we found that the XGBoost model's results remained the same, while the MLP model's performance significantly improved. The inclusion of sentiment scores reduced the MLP model's errors from 7 to 2. This aligns with previous research \cite{mcelfresh2023neural}, which suggests that XGBoost performs better on "irregular" datasets. Our data features indeed exhibit dissimilar distributions, and XGBoost, unlike other models, has the ability to fully utilize the meaning of each data feature.

In the 2020 election, Biden won Arizona and Georgia by very narrow margins of 0.3\% and 0.2\% respectively, making them the two states with the closest victory margins that year. Scholars have pointed out that demographic changes are causing traditionally Republican-leaning states like Arizona and Georgia to become swing states \cite{mclean2018presidential}. However, the XGBoost model may not have been able to capture such patterns in our census data due to the small sample size. On the other hand, the MLP model is relatively more dependent on the trends present in our dataset. For example, the average polling results slightly leaned towards the Democratic Party in 2012 and 2016, and the polling margins in Alaska showed a narrowing trend, accompanied by increasing negative sentiment scores for the Republican Party. To conclude, our findings indicate the potential benefits of leveraging qualitative political analysis as a guide in devising U.S. Election Prediction models.

\begin{table}
\centering
\caption{Evaluation of Classification Models in 2020 U.S. Presidential Election Prediction}
\begin{tabular}{lcc}
\toprule
\textbf{Classification Models} & \textbf{Accuracy} & \textbf{Elector Votes}\\
\midrule
Gradient Boosting Trees & \textbf{96\%} & DNC(279) GOP(259)\\
Logistic Regression & 90\% & DNC(397) GOP(141)\\
Support Vector Machine & 86\% & DNC(372) GOP(166)\\
Random Forest & 76\% & DNC(438) GOP(100)\\
Gaussian Naive Bayes & 88\% & DNC(320) GOP(218)\\
Multi-Layer Perceptron & \textbf{96\%} & DNC(338) GOP(200)\\
Voting Classifier & 94\% & DNC(353) GOP(185)\\
\bottomrule
\multicolumn{3}{p{7.5cm}}{Note: The true Elector votes of the 2020 U.S. Presidential Election is 306 for Democrat and 232 for Republican.}
\end{tabular}
\label{2020eval}
\end{table}

\begin{table}
\centering
\caption{Classification Report of Models in 2020 U.S. Presidential Election Prediction}
\begin{tabular}{lccccc}
\toprule
\textbf{Models} & \textbf{Threshold} &\textbf{Label} & \textbf{Precision} & \textbf{Recall} & \textbf{F1} \\
\midrule
XGBoost & 90\% & DNC & 1 & 0.92 & 0.96\\
        &     & GOP & 0.93 & 1 & 0.96\\
        
        & 50\% & DNC & 1 & 0.77 & 0.87\\
        &     & GOP & 0.81 & 1 & 0.89\\
\midrule
MLP & 90\% & DNC & 0.93 & 1 & 0.96\\
    &     & GOP & 1 & 0.92 & 0.96\\

    & 50\% & DNC & 0.96 & 0.85 & 0.9\\
    &      & GOP & 0.86 & 0.96 & 0.91\\
\bottomrule
\end{tabular}
\label{XGB&FNN}
\end{table}

\section{Discussions and Conclusion}
This study represents a contribution to the field of election prediction, building upon prior research in this area. The polling data's low response and biases could be fixed by applying instant reactions revealed by tweets, as shown in the results of the MLP model. It's important to note that the single polling data is not accurate nowadays due to poor demographic representation and social desirability bias \cite{zhou2021polls}. In our work, we prove that constructing a comprehensive dataset enhances overall accuracy, and thus having a more accurate prediction result compared to previous forecasts \cite{nollenberger2020fundamentals}. 

This study has some limitations, especially for data collection. Some previous polling data and census data are hard to find especially at the state level, considering that the U.S. Presidential Election is held every four years. The data size is reduced because of geographical location. Besides, regions with low populations are underrepresented on X, especially in some swing states. Due to the increasing limitations on the use of third-party APIs in popular social media platforms like X and Reddit, gathering specific data has become more challenging. In upcoming research endeavors, scholars can consider incorporating platforms such as YouTube or TikTok to examine social opinions. In this regard, Lima et al. \cite{lima2023use} demonstrate that election prediction could benefit from a multi-platform approach while investigating the relationship between performance on platforms like TikTok and election outcomes. Another possible direction is to utilize Twitter firehose data for better coverage.

A notable challenge in utilizing social media data, as highlighted by Zhao and Jin \cite{ZhaoJin2020PSPo}, is the presence of sarcasm and humor in user-generated content. These forms of expression can sometimes be misinterpreted, leading to positive sentiment being mistakenly identified as negative, and vice versa. While the Robustly optimized BERT approach (RoBERTa) model is well-structured, there are advanced methods like GPT-4 that offers a more context-aware analysis of user-generated content, thereby improving the accuracy of sentiment analysis. Furthermore, the development of a training set specifically focused on tweets could be beneficial for fine-tuning these models to recognize the political stance of tweets, rather than just sentiments.

Overall, our proposed methodology provides a promising trajectory for electoral forecasting, and our case study on the 2020 U.S. Presidential Election serves as a valuable exemplification of the potential of this approach. Our threshold tuning technique could be applied based on the level of confidence in predicting states. Our methods can serve as a reference for election prediction in other countries, and future research can further refine and expand this methodology, thereby enhancing the accuracy and dependability of electoral predictions.

\section*{Acknowledgment}
This work is supported in part by the Guangdong Provincial Key Laboratory of Interdisciplinary Research and Application for Data Science, BNU-HKBU
United International College, project code 2022B1212010006 and UIC research
grant R0400001-22, and UIC Start-up Research Fund/Grant UICR0700024-22.

\bibliographystyle{IEEEtran}
\bibliography{resources/citation}

\begin{thebibliography}{10}
\providecommand{\url}[1]{#1}
\csname url@samestyle\endcsname
\providecommand{\newblock}{\relax}
\providecommand{\bibinfo}[2]{#2}
\providecommand{\BIBentrySTDinterwordspacing}{\spaceskip=0pt\relax}
\providecommand{\BIBentryALTinterwordstretchfactor}{4}
\providecommand{\BIBentryALTinterwordspacing}{\spaceskip=\fontdimen2\font plus
\BIBentryALTinterwordstretchfactor\fontdimen3\font minus \fontdimen4\font\relax}
\providecommand{\BIBforeignlanguage}[2]{{%
\expandafter\ifx\csname l@#1\endcsname\relax
\typeout{** WARNING: IEEEtran.bst: No hyphenation pattern has been}%
\typeout{** loaded for the language `#1'. Using the pattern for}%
\typeout{** the default language instead.}%
\else
\language=\csname l@#1\endcsname
\fi
#2}}
\providecommand{\BIBdecl}{\relax}
\BIBdecl

\bibitem{fair2011predicting}
R.~Fair, \emph{Predicting presidential elections and other things}.\hskip 1em plus 0.5em minus 0.4em\relax Stanford University Press, 2011.

\bibitem{LewisBeck2014USPE}
\BIBentryALTinterwordspacing
M.~S. Lewis-Beck and M.~Stegmaier, ``Us presidential election forecasting,'' \emph{PS: Political Science \& Politics}, vol.~47, pp. 284 -- 288, 2014. [Online]. Available: \url{https://api.semanticscholar.org/CorpusID:154389407}
\BIBentrySTDinterwordspacing

\bibitem{silver2012signal}
N.~Silver, \emph{The signal and the noise: the art and science of prediction}.\hskip 1em plus 0.5em minus 0.4em\relax Penguin UK, 2012.

\bibitem{linzer2013dynamic}
D.~A. Linzer, ``Dynamic bayesian forecasting of presidential elections in the states,'' \emph{Journal of the American Statistical Association}, vol. 108, no. 501, pp. 124--134, 2013.

\bibitem{jackman2014predictive}
S.~Jackman, ``The predictive power of uniform swing,'' \emph{PS: Political Science \& Politics}, vol.~47, no.~2, pp. 317--321, 2014.

\bibitem{Campbell2017ARO}
\BIBentryALTinterwordspacing
J.~E. Campbell, H.~Norpoth, A.~I. Abramowitz, M.~S. Lewis-Beck, C.~Tien, R.~S. Erikson, C.~Wlezien, B.~Lockerbie, T.~M. Holbrook, B.~J{\'e}r{\^o}me, V.~J{\'e}r{\^o}me-Speziari, A.~Graefe, J.~S. Armstrong, R.~J. Jones, and A.~G. Cuz{\'a}n, ``A recap of the 2016 election forecasts,'' \emph{PS: Political Science \& Politics}, vol.~50, pp. 331 -- 338, 2017. [Online]. Available: \url{https://api.semanticscholar.org/CorpusID:157322250}
\BIBentrySTDinterwordspacing

\bibitem{Valentino2017PollingAP}
\BIBentryALTinterwordspacing
N.~A. Valentino, J.~L. King, and W.~W. Hill, ``Polling and prediction in the 2016 presidential election,'' \emph{Computer}, vol.~50, pp. 110--115, 2017. [Online]. Available: \url{https://api.semanticscholar.org/CorpusID:28916258}
\BIBentrySTDinterwordspacing

\bibitem{OConnor2010FromTT}
B.~O'Connor, R.~Balasubramanyan, B.~Routledge, and N.~Smith, ``From tweets to polls: Linking text sentiment to public opinion time series,'' in \emph{Proceedings of the international AAAI conference on web and social media}, vol.~4, no.~1, 2010, pp. 122--129.

\bibitem{Beauchamp2017PredictingAI}
\BIBentryALTinterwordspacing
N.~Beauchamp, ``Predicting and interpolating state‐level polls using twitter textual data,'' \emph{American Journal of Political Science}, vol.~61, pp. 490--503, 2017. [Online]. Available: \url{https://api.semanticscholar.org/CorpusID:13150074}
\BIBentrySTDinterwordspacing

\bibitem{Yavari2022ElectionPB}
\BIBentryALTinterwordspacing
A.~Yavari, H.~Hassanpour, B.~R. Cami, and M.~Mahdavi, ``Election prediction based on sentiment analysis using twitter data,'' \emph{International Journal of Engineering}, 2022. [Online]. Available: \url{https://api.semanticscholar.org/CorpusID:245347302}
\BIBentrySTDinterwordspacing

\bibitem{Mislove2011UnderstandingTD}
\BIBentryALTinterwordspacing
A.~Mislove, S.~Lehmann, Y.-Y. Ahn, J.-P. Onnela, and J.~N. Rosenquist, ``Understanding the demographics of twitter users,'' \emph{Proceedings of the International AAAI Conference on Web and Social Media}, 2011. [Online]. Available: \url{https://api.semanticscholar.org/CorpusID:15924076}
\BIBentrySTDinterwordspacing

\bibitem{Anuta2017ElectionBC}
\BIBentryALTinterwordspacing
D.~Anuta, J.~Churchin, and J.~Luo, ``Election bias: Comparing polls and twitter in the 2016 u.s. election,'' \emph{ArXiv}, vol. abs/1701.06232, 2017. [Online]. Available: \url{https://api.semanticscholar.org/CorpusID:18048854}
\BIBentrySTDinterwordspacing

\bibitem{liu2021can}
R.~Liu, X.~Yao, C.~Guo, and X.~Wei, ``Can we forecast presidential election using twitter data? an integrative modelling approach,'' \emph{Annals of GIS}, vol.~27, no.~1, pp. 43--56, 2021.

\bibitem{CensusReference}
\BIBentryALTinterwordspacing
D.~W. Nickerson and T.~Rogers, ``Political campaigns and big data,'' \emph{Journal of Economic Perspectives}, vol.~28, no.~2, pp. 51--74, May 2014. [Online]. Available: \url{https://www.aeaweb.org/articles?id=10.1257/jep.28.2.51}
\BIBentrySTDinterwordspacing

\bibitem{houle2019religion}
C.~Houle, ``Religion, language, race and ethnic voting,'' \emph{Electoral Studies}, vol.~61, p. 102052, 2019.

\bibitem{lewis2000economic}
M.~S. Lewis-Beck and M.~Stegmaier, ``Economic determinants of electoral outcomes,'' \emph{Annual review of political science}, vol.~3, no.~1, pp. 183--219, 2000.

\bibitem{TurnerMcGrievy2015TweetFH}
G.~M. Turner-McGrievy and M.~W. Beets, ``Tweet for health: using an online social network to examine temporal trends in weight loss-related posts,'' \emph{Translational Behavioral Medicine}, vol.~5, pp. 160--166, 2015.

\bibitem{Harris2014ArePH}
J.~K. Harris, B.~Choucair, R.~C. Maier, N.~Jolani, and J.~M. Bernhardt, ``Are public health organizations tweeting to the choir? understanding local health department twitter followership,'' \emph{Journal of Medical Internet Research}, vol.~16, 2014.

\bibitem{loureiro-etal-2022-timelms}
\BIBentryALTinterwordspacing
D.~Loureiro, F.~Barbieri, L.~Neves, L.~Espinosa~Anke, and J.~Camacho-collados, ``{T}ime{LM}s: Diachronic language models from {T}witter,'' in \emph{Proceedings of the 60th Annual Meeting of the Association for Computational Linguistics: System Demonstrations}.\hskip 1em plus 0.5em minus 0.4em\relax Dublin, Ireland: Association for Computational Linguistics, May 2022, pp. 251--260. [Online]. Available: \url{https://aclanthology.org/2022.acl-demo.25}
\BIBentrySTDinterwordspacing

\bibitem{Liu2019RoBERTaAR}
\BIBentryALTinterwordspacing
Y.~Liu, M.~Ott, N.~Goyal, J.~Du, M.~Joshi, D.~Chen, O.~Levy, M.~Lewis, L.~Zettlemoyer, and V.~Stoyanov, ``Roberta: A robustly optimized bert pretraining approach,'' \emph{ArXiv}, vol. abs/1907.11692, 2019. [Online]. Available: \url{https://api.semanticscholar.org/CorpusID:198953378}
\BIBentrySTDinterwordspacing

\bibitem{shwartz2022tabular}
R.~Shwartz-Ziv and A.~Armon, ``Tabular data: Deep learning is not all you need,'' \emph{Information Fusion}, vol.~81, pp. 84--90, 2022.

\bibitem{mcelfresh2023neural}
D.~McElfresh, S.~Khandagale, J.~Valverde, G.~Ramakrishnan, M.~Goldblum, C.~White \emph{et~al.}, ``When do neural nets outperform boosted trees on tabular data?'' \emph{arXiv preprint arXiv:2305.02997}, 2023.

\bibitem{mclean2018presidential}
S.~L. McLean, S.~D. Foreman, D.~R. Hoffman, C.~W. Larimer, D.~J. Scala, D.~F. Damore, R.~D. Gill, S.~Trende, R.~R. Preuhs, N.~Provizer \emph{et~al.}, \emph{Presidential Swing States}.\hskip 1em plus 0.5em minus 0.4em\relax Rowman \& Littlefield, 2018.

\bibitem{zhou2021polls}
Z.~Zhou, M.~Serafino, L.~Cohan, G.~Caldarelli, and H.~A. Makse, ``Why polls fail to predict elections,'' 2021.

\bibitem{nollenberger2020fundamentals}
C.~Nollenberger and G.-M. Unger, ``Fundamentals-based state-level forecasts of the 2020 us presidential election,'' 2020.

\bibitem{lima2023use}
J.~Lima, M.~Santana, A.~Correa, and K.~Brito, ``The use and impact of tiktok in the 2022 brazilian presidential election,'' in \emph{Proceedings of the 24th Annual International Conference on Digital Government Research}, 2023, pp. 144--152.

\bibitem{ZhaoJin2020PSPo}
J.~Zhao, ``\BIBforeignlanguage{eng}{Political stance prediction on youtube comments},'' 2020.

\end{thebibliography}

\end{document}